\documentclass[%
 reprint,
 amsmath,amssymb,
 aps,
]{revtex4-2}

\usepackage{graphicx}
\usepackage{dcolumn}
\usepackage{bm}
\usepackage{textcomp,mathcomp}
\usepackage{verbatim}
\usepackage{caption}
\usepackage{subcaption}

\begin{document}

\preprint{APS/123-QED}

\title{Designing Pu Intermetallics with First Principle Calculations}

\author{Matthew S. Cook\textsuperscript{†}}
\author{David C. Arellano\textsuperscript{†}}
\author{Derek V. Prada\textsuperscript{†}}
\author{Sven P. Rudin\textsuperscript{††}}
\author{Eric D. Bauer\textsuperscript{†††}}
\author{W. Adam Phelan\textsuperscript{†}*}

\affiliation{%
 \textsuperscript{†}MST-16, Los Alamos National Laboratory, Los Alamos, New Mexico 87545, USA}
 \affiliation{%
 \textsuperscript{††}T-1, Los Alamos National Laboratory, Los Alamos, New Mexico 87545, USA}
 \affiliation{%
 \textsuperscript{†††}MPA-Q, Los Alamos National Laboratory, Los Alamos, New Mexico 87545, USA}
 \affiliation{%
 *corresponding author}

\date{\today}

\begin{abstract} 
We present the \textit{ab initio} supported discovery of two new Pu based intermetallic compounds, PuNiSn and PuPtSn. Using density functional theory, the formation energies within the relevant ternary phase diagrams were calculated to predict the stability of both compounds. Simultaneously, Pu-Ni-Sn and Pu-Pt-Sn materials were arc-melted and subsequently characterized with magnetization, specific heat, and resistivity measurements from 2-300 K. Magnetization measurements show that PuNiSn and PuPtSn order antiferromagnetically at $T_N=11\ K$ and $T_N=15\ K$, respectively. Specific heat measurements show an enhanced residual electronic specific heat that is indicative of strong electron correlations. Resistivity measurements are indicative of Kondo behavior for PuNiSn while crystal field effects may play a role in the observed temperature dependence for PuPtSn.   

\end{abstract}

\maketitle


\section{\label{sec:level1}Introduction\protect\\}
Plutonium (Pu) is a unique element with a complex phase diagram that hosts seven allotropic phases \cite{troc_bulk_1985,hecker_magic_2004}, an electronically driven 25\% volume contraction from the $\delta$ phase to the low temperature $\alpha$-phase \cite{harrison_phase_2019}, and an anomalously large residual electronic heat capacity has been observed for its meta-stable alloy phases (e.g., Ga-stabilized $\delta$-Pu) \cite{lashley_experimental_2003}. Many of theses extraordinary properties of Pu are generally associated with the unstable nature of its 5f electron configuration, which resides at the border between itinerant and localized behavior in the periodic table. 

This complexity is also clearly expressed in Pu compounds, where the 5f-states can vary from 5f\textsuperscript{2} to 5f\textsuperscript{6} \cite{noauthor_chemistry_2006}, or non-integer filling \cite{joyce_dual_2006,shick_electronic-structure_2009,ramshaw_avoided_2015}. In this regard, Pu-based intermetallic compounds have attracted interest for their complex physical behavior which results from strong hybridization of 5f-states with the conduction electron states of the surrounding ligands.  The strong 5f correlations in Pu intermetallics often result in compounds with an enhanced electronic specific heat coefficient, which corresponds to a large effective mass of charge carriers \cite{bauer_plutonium-based_2015}. Various magnetic ground states have also been observed in Pu compounds, such as ferromagnetism \cite{griveau_magnetic_2013}, antiferromagnetism \cite{bauer_single_2012}, and paramagnetism \cite{wastin_magnetic_1993}. The four known examples of Pu-based superconductors, PuCoGa$_5$, PuRhGa$_5$, PuCoIn$_5$, and PuRhIn$_5$ are characterized by enhanced electronic specific heat coefficients and relatively large superconducting transition temperatures for this superconducting class \cite{sarrao_plutonium-based_2002, wastin_advances_2003, bauer_localized_2012,bauer_single_2012}. For heavy fermion superconductors, PuCoGa\textsubscript{5} was found to have a record high transition temperature of 18.5 K, which is several times larger than isostructural Ce based counterparts, CeCoIn\textsubscript{5} and CeIrIn\textsubscript{5} \cite{petrovic_heavy-fermion_2001,d_thompson_progress_2012}.

From the preceding statements, the discovery of new Pu-containing materials is a worthy pursuit. Yet, the experimental search for new Pu-based intermetallics is heavily constrained by the acquisition of starting material, its associated safety hazards concerning radioactivity, and the limited laboratory spaces equipped to handled Pu. On the other hand, the theoretical search for such intermetallics is limited by the challenges that the strong 5f correlations pose on methods. Thus, it is vital that experimentalists and theorists work together to make informed predictions on the stability of potential new phases. In this work, by using density functional theory (DFT) to calculate thermodynamics parameters of various ternary and binary Pu intermetallic compounds in the Pu:T:Sn (T = transition metals) phase space, we attempt to marry both experiment and theory in the hopes of laying ground for future theory driven Pu-materials discovery. 

In addition to the aforementioned DFT component, we report the synthesis and physical properties characterization of two new Pu-based intermetallics, namely, PuNiSn and PuPtSn. These materials expand the phase space of Pu 1:1:1 intermetallic compounds, as similar PuTX (T = transition metals and X = Al, Ga) compounds were first synthesized nearly 5 decades ago \cite{sechovsky_chapter_1988}. More recently, the physical characterization of PuPdSn was reported by Gofryk et al. \cite{gofryk_magnetic_2009}. Notably, PuNiSn crystallizes in an orthorhombic TiNiSi-type structure (space group $Pnma$) while PuPtSn shares the same structure as the Pd analog. Further, using magnetization we find that PuNiSn orders antiferromagnetically at $T_N=11\ $K while PuPtSn also shows an antiferromagnetic transition at $T_N=15\ $K with a potential secondary transition at 3 K. Curie-Weiss fits to the high temperature magnetic susceptibility show that the effective moment for PuNiSn ($\mu_{eff}=0.93\ \mu_B$) is close to that expected for an intermediate coupling scheme, while PuPtSn ($\mu_{eff}=0.77\ \mu_B$) is slightly lower than that expected in a Russell-Saunders coupling scheme. Low temperature fits to $C/T$ vs $T^2$ taken above the magnetic transitions find a Sommerfeld coefficient $\gamma$ of $243(3)\ \mathrm{mJ}\ \mathrm{mol}^{-1}\ \mathrm{K}^{-2}$ for PuNiSn and $216(3)\ \mathrm{mJ}\ \mathrm{mol}^{-1}\ \mathrm{K}^{-2}$ for PuPtSn. Resistivity measurements show a weak temperature dependence in both materials with clear Kondo behavior present in PuNiSn. Finally, using DFT, we calculate the phonon dispersion and corresponding partial density of states for both compounds and compare to the lattice modes determined from the specific heat. We find that two Debye modes can effectively capture the vibrational frequencies coming from the consituent atoms due to their large difference in mass. 

\section{Methods}

DFT calculations were performed to determine total energies and phonon dispersions/density of states for a variety of materials within the Pu-Ni-Sn and Pu-Pt-Sn phase diagrams. The DFT calculations allowed (collinear) spin polarization (orbital polarization is not included). The computational cells contained multiple Pu atoms, and so all possible spin structures were tested. Here we allow spin polarization because the ground state of Pu compounds often have a net magnetic moment. The inclusion of spin polarization has also been shown to result in a more accurate description of elemental Pu in approximation of the strong correlations present in the 5f electronic structure \cite{soderlind_density-functional_2019}. For the two ternary compounds reported here, the spins on all Pu atoms aligned in the same direction (ferromagnet). Spin-orbit coupling is neglected, as its effects on the results for representative structures show negligible changes. The DFT calculations were performed using the VASP package \cite{kresse_efficient_1996,kresse_ultrasoft_1999}. The electronic states were treated with the projector augmented wave method \cite{blochl_projector_1994}, in the generalized gradient approximation of Perdew, Burke, and Ernzerhof \cite{perdew_generalized_1996}, with first-order Methfessel-Paxton smearing (width 27 meV) \cite{methfessel_high-precision_1989}, and a cutoff energy for the plane wave basis set of 500 eV. The k-point meshes were chosen for each simulation cell such that their density is at least 60 per \AA\textsuperscript{-1}. The convergence criteria are 10\textsuperscript{-7} eV for the electronic self-consistency loop and 10\textsuperscript{-6} eV for the ionic relaxation. The phonon calculations employ 2x2x2 supercells. Ternary phase diagrams were explored by calculating formation energies.\cite{fredeman_computationally_2011}
Since the focus is on the central part of the phase diagrams, the complete convex hull is not calculated. Instead, formation energies are calculated relative to known, nearby compounds.

\begin{figure}[!ht]
\includegraphics[width=3.375in]{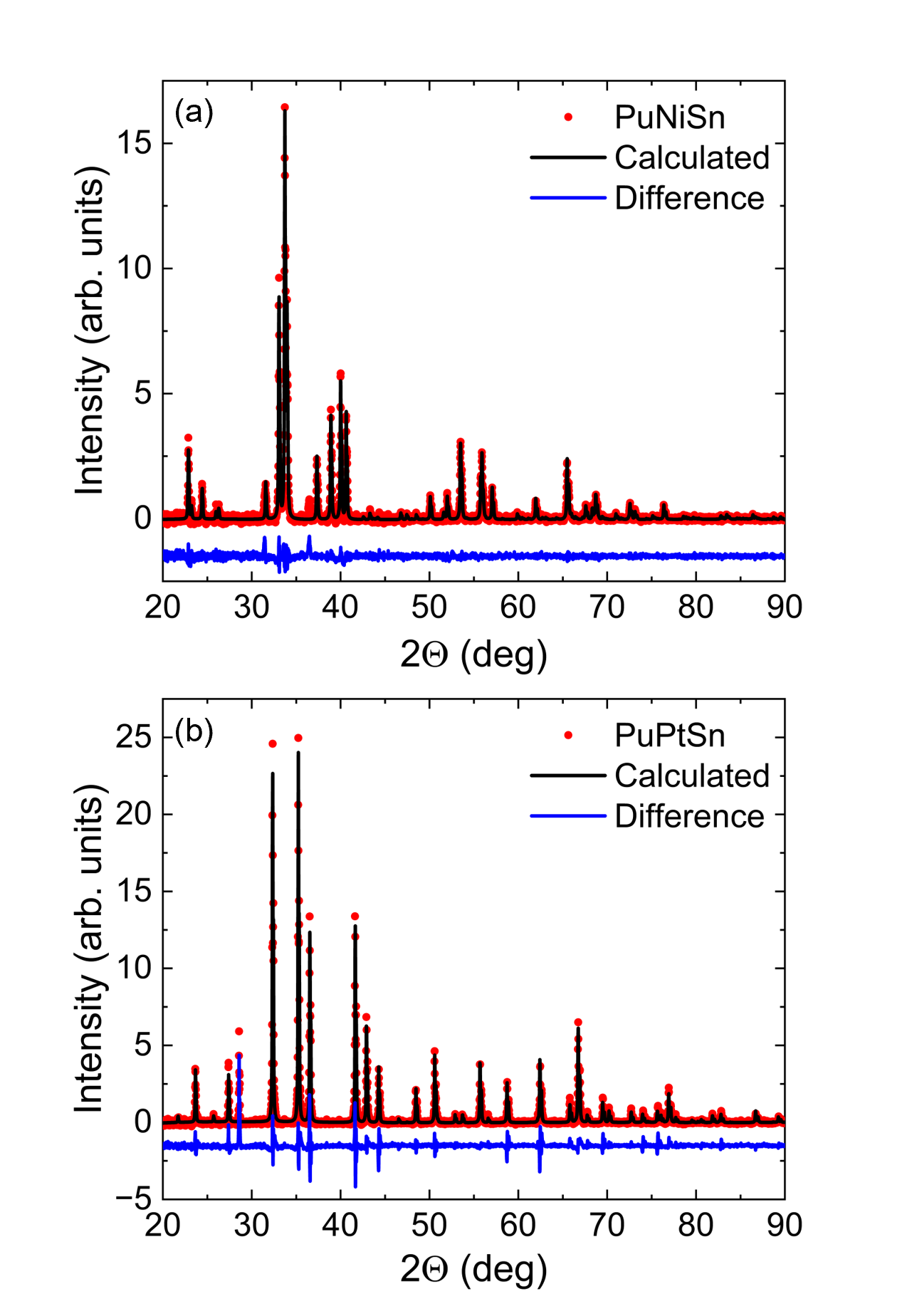}
\caption{\label{fig:xrd} (a) Powder x-ray diffraction data (red circles) for PuNiSn with LeBail fit overlain in black. (b) Powder x-ray diffraction data (red circles) for PuPtSn with LeBail fit overlain in black. The difference between the calculated and experimental data are shown in blue in both panels.}
\end{figure}

Polycrystalline samples of PuNiSn and PuPtSn were arc-melted on stoichiometry in an Zr-gettered argon atmosphere. Each button was flipped and remelted twice to help ensure homogenization. Following arc-melting, the buttons were wrapped in Ta-foil and sealed into separate evacuated silica tubes. The tubes were placed into a muffle furnace at 773 K for two weeks.

Powder X-ray diffraction data were collected at room temperature using a Bruker D8 Discover with DAVINCI design powder diffractometer equipped with a LynxEye detector and a Cu K$\alpha$ radiation source. Subsequent LeBail fits were generated using Bruker’s TOPAS v. 6 sofware. PuNiSn crystallizes in an orthorhombic TiNiSi-type structure (space group $Pnma$), as shown in Fig.~\ref{fig:xrd}(a). PuPtSn crystallizes in a hexagonal ZrNiAl-type structure (space group $P\overline{6}2m$), as shown in Fig.~\ref{fig:xrd}(b).

Specific heat measurements were collected using a commercial physical properties measurement system (PPMS) cryostat via a semi-adiabatic thermal relaxation technique. As Pu is a radioactive element, decay in this material causes self-heating. This limits the temperature of which we were able to cool the sample, such that larger samples were more difficult to cool. For PuNiSn, the mass of the measured sample is 9 mg, thus we were only able to measure to approximately $T=5.5\ \mathrm{K}$. For a sample mass of only 1.7 mg for PuPtSn, this allowed us to reach $T=2.5\ \mathrm{K}$. Magnetization measurements were also performed in the same commercial PPMS cryostat using a vibrating sample magnetometer. In order to safely measure Pu samples and to prevent possible contamination hazards, the magnetization samples were mounted using  layers of encapsulation. This includes small amounts of plastic wrap and GE varnish glued inside a plastic straw. The plastic straw was then sealed with custom caps that contain micron sized platinum frits which were glued to the straw with stycast epoxy. Resistivity measurements were done with a standard 4-wire configuration in the PPMS Dynacool cryostat using native electronics.


\section{Results and Discussion}

\subsection{Convex Hull Calculations}

To identify potential candidates for Pu-materials discovery, we determine the theoretical thermodynamic stability of various compounds in the Pu-Ni-Sn and Pu-Pt-Sn phase spaces using formation energy calculations and a convex hull construction \cite{fredeman_computationally_2011}. We calculate the formation energy of a given compound of composition Pu$_l$T$_m$Sn$_n$ relative to its constituent elements as:
\begin{eqnarray}
\Delta{E_f}={E_{\mathrm{Pu}_l\mathrm{T}_m\mathrm{Sn}_n}-lE_{\mathrm{Pu}}-mE_{\mathrm{T}}-nE_{\mathrm{Sn}}},
\label{eq:DChall}
\end{eqnarray}
where $E_{\mathrm{Pu}_l\mathrm{T}_m\mathrm{Sn}_n}$ is the total energy per formula unit of $\mathrm{Pu}_l\mathrm{T}_m\mathrm{Sn}_n$, while $E_{\mathrm{Pu}}$, $E_\mathrm{T}$, and $E_{\mathrm{Sn}}$ are the total energies per atom of the constituent elements.

\begin{figure}[h]
\includegraphics[width=3.375in]{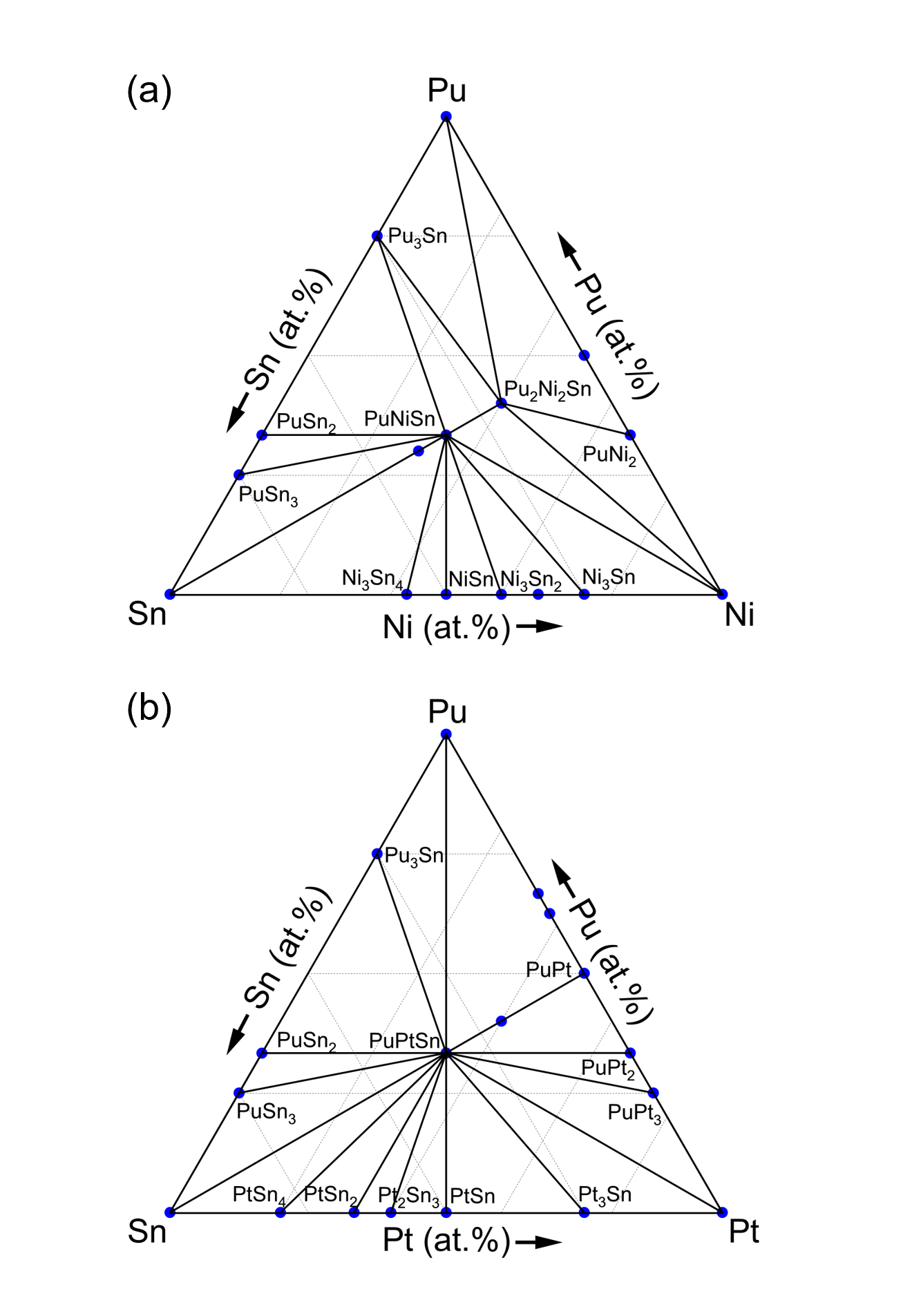}
\caption{\label{fig:convex} DFT predicted phase diagrams for the (a) Pu-Ni-Sn and (b) Pu-Pt-Sn systems. The phases which fall on a vertex of the facets are expected to be thermodynamically stable compounds.}
\end{figure}

In Fig.~\ref{fig:convex}, we plot each phase for the (a) Pu-Ni-Sn and (b) Pu-Pt-Sn series at its particular composition in a two-dimensional diagram. We then form both convex hulls using the calculated formation energies of each composition relative to their constituent elements. This allows for identification of competing phases in a relative compositional space. We also calculate the formation energy ($E_{comp}$) with respect to these competing phases to determine if decomposition is energetically favorable and ultimately determine a compound's theoretical thermodynamic stability at zero temperature. The predicted stable compounds from this process lie on the vertex of each facet in Fig.~~\ref{fig:convex}. Our calculations show that both PuNiSn and PuPtSn are predicted to be thermodynamically stable relative to nearby competing phases. Table 1 presents the complete results of this analysis for both compounds. 

\begin{table*}[ht]
    \centering
    \begin{tabular}{cccccc} 
         &  $E_{int}$ &  $\Delta{E_f}$ &  $\Delta{E_{comp}}$ &  & \\
 Compound& [eV/f.u.]& [meV/atom]& [meV/atom]& Competing Phases&Space Group\\ \hline
         PuNiSn&  -24.377&  -343&  -67&  Pu$_2$Ni$_2$Sn, Pu$_3$Ni$_3$Sn$_4$& $Pnma$ (62)\\ 
         Pu$_2$Ni$_2$Sn&  -44.339&  296&  -4&  PuNiSn, PuNi$_2$, Pu& $P4/mbm$ (127)\\ 
         Pu$_3$Ni$_3$Sn$_4$&  -76.542&  -266&  40&  PuNiSn, Sn& $I\bar{4}3D$ (220)\\ 
         Pu$_3$Sn&  -46.708&  -185&  -49&  Pu, PuSn$_2$& $Pm\bar{3}m$ (221)\\ 
         PuSn$_2$&  -22.801&  -363&  -185&  Pu$_3$Sn, PuSn$_3$& $I4_1/amd$ (141)\\ 
         PuSn$_3$&  -26.946&  -350&  -77&  PuSn$_2$, Sn& $	
Pm\bar{3}m$ (221)\\ 
         PuNi$_2$&  -25.837&  -286&  -217&  Ni, PuNi& $Fd\bar3m$ (227)\\ 
         PuNi&  -19.719&  -103&  111&  Pu, PuNi$_2$& $Cmcm$ (63)\\ 
         Ni$_3$Sn&  -21.004&  -192&  -22&  Ni$_3$Sn$_2$, Ni& 	
$Pm\bar3m$ (221)\\ 
 Ni$_2$Sn& -15.269& -167& 69& Ni$_3$Sn, Ni$_3$Sn$_2$&\\
 Ni$_3$Sn$_2$& -25.427& -271& -38& Ni$_3$Sn, NiSn&$Pnma$ (62)\\
 NiSn& -9.822& -260& -11& Ni$_3$Sn$_2$, Ni$_3$Sn$_4$&$Pbam$ (55)\\
 Ni$_3$Sn$_4$& -33.374& -234& -11& NiSn, Sn&$C2/m$ (12)\\ \hline
 & & & & &\\ \hline
 PuPtSn& -26.187& -738& -169& Pu$_2$Pt$_2$Sn, PtSn, PuSn$_2$, PuPt&$P \bar62m$ (189)\\
 Pu$_2$Pt$_2$Sn& -47.245& -627& 126& PuPtSn, PuPt&$P4/mbm$ (127)\\
 Pu$_2$Pt& -35.629& -480& 34& Pu, PuPt&$Pnma$ (62)\\
 Pu$_5$Pt$_3$& -93.018& -562& 16.2& Pu,PuPt&$P63/mcm$ (193)\\
 PuPt& -21.686& -771& -217& Pu,PuPt$_2$&$Cmcm$ (63)\\
 PuPt$_2$& -28.459& -738& -26& PuPt, PuPt$_3$&$Fd\bar3m$ (227)\\
 PuPt$_3$& -35.074& -683& -129& PuPt$_2$, Pt&$Pm\bar3m$ (221)\\
 Pu$_3$Sn& -46.708& -185& -49& Pu, PuSn$_2$&$Pm\bar3m$ (221)\\
 PuSn$_2$& -22.801& -363& -185& Pu$_3$Sn,PuSn$_3$,Sn&$I4_1/amd$ (141)\\
 PuSn$_3$& -26.946& -350& -77& PuSn$_2$, Sn&$Pm\bar3m$ (221)\\
 Pt$_3$Sn& -23.662& -383& -90& Pt, PtSn&$Pm\bar3m$ (221)\\
 PtSn& -11.102& -585& -129& Pt$_3$Sn, Pt2Sn$_3$, PtSn$_2$&$P6_3/mmc$ (194)\\
 Pt$_2$Sn$_3$& -26.268& -514& -6& PtSn, PtSn$_2$&$P6_3/mmc$ (194)\\
 PtSn$_2$& -15.133& -456& -18& Pt$_2$Sn$_3$, PtSn$_4$&$Fm\bar3m$ (221)\\ 
    \end{tabular}
    \caption{Relevant phases with associated space groups and calculated energies for Pu-Ni-Sn and Pu-Pt-Sn phase diagrams. Determined competing phases for which the formation energy $\Delta{E_{comp}}$ is calculated are also listed.}   
    \label{tab:my_label}
\end{table*}

\subsection{Magnetization}

\begin{figure*}[!ht]
    \begin{center}
    \includegraphics[width=\textwidth]{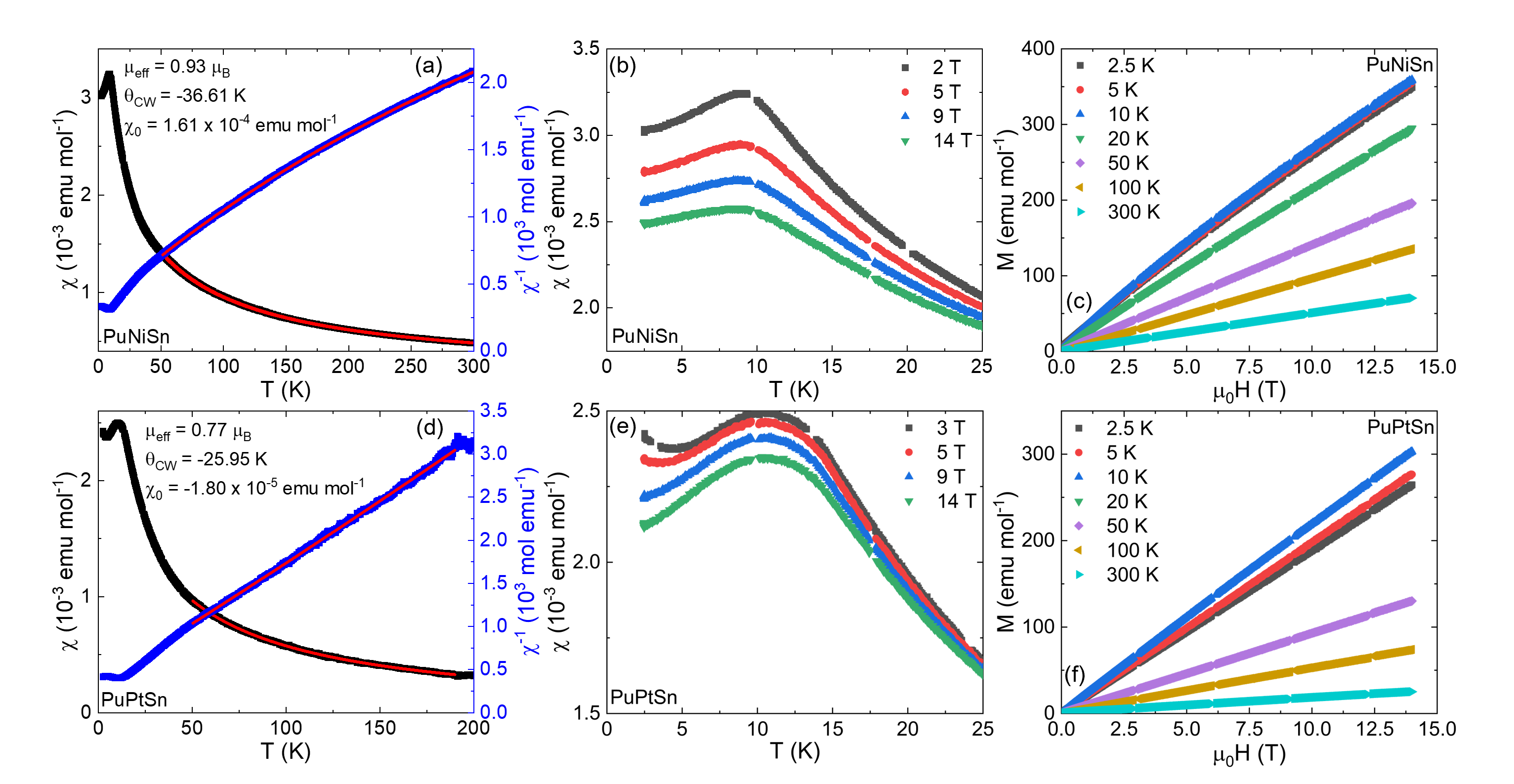}
    \end{center}
    \caption{\label{fig:mag} The magnetic susceptibility $\chi$ (left axis) and corresponding inverse magnetic susceptiblity $\chi^{-1}$ (right axis) for (a) PuNiSn and (d) PuPtSn. The solid red lines correspond to Curie-Weiss fits to the data. Antiferromagnetic transitions in the low temperature magnetic susceptibility $\chi$ at various magnetic fields for (b) PuNiSn and (e) PuPtSn. There is a secondary transition near 4 K for PuPtSn in a field of 3 T. The field dependence of the magnetization $M$ at various temperatures for (c) PuNiSn and (f) PuPtSn.} 
\end{figure*}

The magnetic susceptibility $\chi=M/H$ and its inverse $\chi^{-1}$ are shown in Figs~\ref{fig:mag}(a) and \ref{fig:mag}(d) for PuNiSn and PuPtSn, respectively. Both compounds were fit with the Curie-Weiss law given by:

\begin{eqnarray}
\chi(T)=\chi_0+\frac{C}{T-\theta_C}, 
\label{eq:vrh}
\end{eqnarray}

where $\chi_0$ is the temperature independent contribution that results from a combination of Pauli paramagnetism, Van Vleck paramagnetism, and the sample holder background, $\theta_C$ is the Curie-Weiss temperature, and $C$ is related to the effective moment $\mu_{eff}=\sqrt{8C}\ \mu_B$. The results of the fit give $\theta_C=-37\ K$, $\mu_{eff}=0.93\ \mu_B$, and $\chi_0=161\times10^{-6}$ emu mol$^{-1}$ for PuNiSn, and $\theta_C=-26\ K$, $\mu_{eff}=0.77\ \mu_B$, and $\chi_0=-1.80\times10^{-5}$ emu mol$^{-1}$ for PuPtSn. 

The effective moment for PuNiSn is close to the value expected for intermediate coupling (1.1 $\mu_B$) and similar to that found for Pu\textsubscript{2}Ni\textsubscript{3}Si\textsubscript{5} \cite{bauer_magnetic_2011}. An intermediate coupling scheme has been previously shown to be appropriate in the description of 5f orbitals for f\textsuperscript{5} Pu compounds \cite{shorikov_magnetic_2005}. The negative $\theta_C=-37\ K$ is indicative of antiferromagnetic correlations that lead to an antiferromagnetic transition that can been seen in magnetic susceptibility at $T_N=11\ K$ in Fig.~\ref{fig:mag}(b). Here, $T_N$ is defined as the maximum in the derivative $d\chi/dT$ (not shown). A magnetic field of 14 T has a very small effect on the transition temperature.

For PuPtSn, the effective moment $\mu_{eff}=0.77\ \mu_B$ is slightly below the value expected in an LS coupling scheme for a $5f^5$ manifold ($0.84 \mu_B$). Although the intermediate coupling scheme is generally accepted for Pu\textsuperscript{3+} compounds due to strong spin orbit coupling, lower values for the effective moment have also been observed in other Pu intermetallics \cite{boulet_magnetic_2005}. The magnetic transition at $T_N=15\ K$ is in agreement with the negative $\theta_{C}$ and also has a negligible dependence on magnetic field, as shown in Fig.~\ref{fig:mag}(e). The AFM transition in PuPtSn is extremely broad compared to PuNiSn and this character is also observed in the specific heat. There is a secondary transition in $\chi$ that is seen as an upturn in $\chi$ vs $T$ at approximately 4 K. This transition is suppressed by the application of a field of 9 T.

The magnetism and strong electron correlations of related uranium ternary compounds with the orthorhombic TiNiSi-type and hexagonal ZrNiAl-type structures have been previously discussed in the context of electron hybridization that is a direct consequence of the large spacial extent of the 5f wavefunction \cite{sechovsky_chapter_1988}. The hybridization of 5f states with valence states of the surrounding atoms in these materials typically results in the delocalization of 5f electrons and the formation of a narrow 5f band that intersects the Fermi energy ($E_f$). This leads to magnetic moments that are much smaller than expected for a free U$^{3+}$ ion as well as a large density of states at $E_f$ that is reflected in an enhanced electronic specific heat coefficient $\gamma$. Here, the effective moment estimated in both compounds is not greatly reduced from the expected moment for a free Pu$^{3+}$ ion, thus we can deduce that the 5f electrons in these materials have a localized character.

\subsection{Specific Heat}

\begin{figure*}[!ht]
    \begin{center}
    \includegraphics[width=\textwidth]{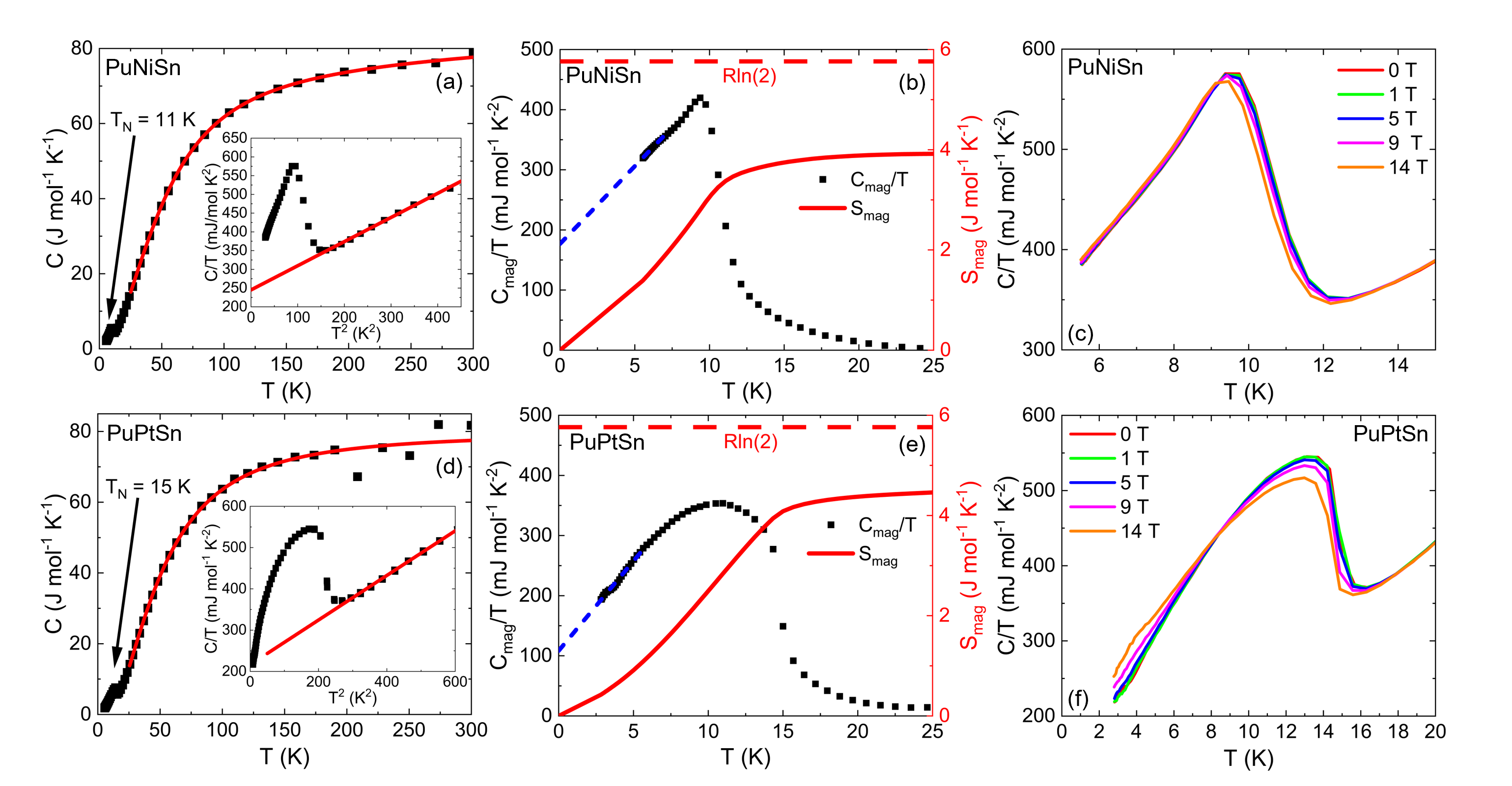}
    \end{center}
    \caption{\label{fig:specific} Specific heat of (a) PuNiSn and (d) PuPtSn with corresponding fits to equation (3) in solid red. Insets: Peak in $C/T$ vs $T^2$ from AFM order with linear fit in solid red. Magnetic specific heat divided by temperature is plotted on the left axis of (b) for PuNiSn and (e) PuPtSn, while the magnetic entropies are shown in red on the right axes. The magnetic field dependence of the specific heat divided by temperature for (c) PuNiSn and (f) PuPtSn.}
\end{figure*}

The zero-field specific heat of PuNiSn and PuPtSn for $5\ \mathrm{K}\le300\ \mathrm{K}$ is shown in Fig.~\ref{fig:specific}(a) and Fig.~\ref{fig:specific}(d), respectively. For PuNiSn, there is an anomaly in the specific heat at $T_N=11\ \mathrm{K}$ that corresponds to the onset of antiferromagentic order. For PuPtSn, the ordering temperature occurs slightly higher, at $T_N=15\ \mathrm{K}$. This is in contrast to the isostructural PuPdSn which shows two magnetic transitions in specific heat and magnetization measurements at $T_{1}=21\ $K and $T_{2}=9.6\ $K \cite{gofryk_magnetic_2009}. To get an estimate of the Sommerfeld coefficient at low temperatures, the specific heat of both compounds is plotted as $C/T$ vs $T^2$, as shown in the inset of Figs~\ref{fig:specific}(a) and (d). We fit the linear region above $T_n$ to $C/T=\gamma+{\beta}T^2$, where $\gamma$ is the Sommerfeld coefficient and $\beta=1944/\Theta_D^3$, where $\Theta_D$ represents the Debye temperature. The fit for PuNiSn yields $\gamma=243(3)\ \mathrm{mJ}\ \mathrm{mol}^{-1}\ \mathrm{K}^{-2}$ and $\beta=0.653(8)\ \mathrm{mJ}\ \mathrm{mol}^{-1}\ \mathrm{K}^{-4}$. Similarly, the fit for PuPtSn yields $\gamma=216(2)\ \mathrm{mJ}\ \mathrm{mol}^{-1}\ \mathrm{K}^{-2}$ and $\beta=0.540(5)\ \mathrm{mJ}\ \mathrm{mol}^{-1}\ \mathrm{K}^{-4}$. The estimated $\gamma$ for both compounds is fairly enhanced when compared with PuPdSn, which is estimated as $8\ \mathrm{mJ}\ \mathrm{mol}^{-1}\ \mathrm{K}^{-2}$ \cite{gofryk_magnetic_2009}. However, the value found for PuPdSn must be met with caution as it was determined by fitting the specific heat below the magnetic transition with an electronic and debye term. To make a more direct comparison to PuPdSn, we digitize the specific heat data in Ref.~\cite{gofryk_magnetic_2009} and determine $\gamma$ via extrapolation to 0 K in the paramagnetic state (not shown). Using this method, we find a much larger $\gamma=290$ mJ mol\textsuperscript{-1} K\textsuperscript{-2}, which is closer to the values found for PuNiSn and PuPtSn. The values estimated for $\gamma$ are also similar to other magnetically ordered plutonium based heavy-fermion systems, such as PuIn\textsubscript{3} ($307\ \mathrm{mJ}\ \mathrm{mol}^{-1}\ \mathrm{K}^{-2}$) \cite{chudo_weak_2014}, PuAl\textsubscript{2} ($260\ \mathrm{mJ}\ \mathrm{mol}^{-1}\ \mathrm{K}^{-2}$) \cite{stewart_anomalous_1985}, and Pu$_2$Co$_3$Si$_5$ ($95\ \mathrm{mJ}\ \mathrm{mol}^{-1}\ \mathrm{K}^{-2}$),
indicative of strong electron correlations.  

In order to determine an accurate contribution to the specific heat from the lattice, we employed a fit from $25\ \mathrm{K}\le300\ \mathrm{K}$. The magnetic transitions at low temperatures causes difficulty in getting a reliable fit over the entire temperature range, thus the chosen temperatures are well above the transition. We assume that the form of the specific heat is given by $C=C_e+(n-r)C_{D1}+(r)C_{D2}$, where $n=3$ is the number of atoms per formula unit, r is a fitting parameter used to keep the amount of oscillators consistent, $C_e=\gamma{T}$ where $\gamma$ is the Sommerfeld coefficient, and $C_{D}$ is a Debye term given by 
\begin{equation}
C_D = 9Nk_Bx^{-3}\int_{0}^{x}\frac{x^4e^x}{(e^x-1)^2} \,dx
\end{equation} 
where $T_D$ is the Debye temperature and $x=T_D/T$. Attempts to fit the specific heat using a single Debye mode were attempted, yet were unsuccessful. The resulting fits that are plotted as the solid red lines in Fig.~\ref{fig:specific}(a) and (d) yield $T_{D1}=259(10)\ \mathrm{K}$, $T_{D2}=113(12)\ \mathrm{K}$, $\gamma=16.2(5)\ \mathrm{mJ}\ \mathrm{mol}^{-1}\ \mathrm{K}^{-2}$, $r=1.0(2)$ for PuNiSn and $T_{D1}=266(11)\ \mathrm{K}$, $T_{D2}=141(8)\ \mathrm{K}$, $\gamma=5.2(5)\ \mathrm{mJ}\ \mathrm{mol}^{-1}\ \mathrm{K}^{-2}$, $r=1.3(2)$ for PuPtSn. We will further comment on the lattice specific heat contributions in the next section in relation to the phonon density of states.

To isolate the magnetic contribution to the specific heat, we subtract the fit to the high temperature specific heat such that $C_{mag}=C-(C_e+(n-r)C_{D1}+(r)C_{D2})$. To get an estimate of the magnetic entropy associated with the AFM order in this system, we tentatively extrapolate $C_{mag}/T$ vs $T$ linearly to $T=0\ \mathrm{K}$, as shown in blue on the left axes in Figs~\ref{fig:specific}(b) and (e). $C_{mag}(T)/T$ vs $T$ is then numerically integrated to calculate the magnetic entropy $S_{mag}(T)$, which we plot on the right axes of Fig.~\ref{fig:specific}(b) and (e). 

For both compounds, it is possible that crystalline electric field (CEF) effects give rise to a doublet ground state; the J = 5/2 multiplet of a Pu$^{3+}$ ion in an orthorhombic or hexagonal site symmetry split into 3 Kramers doublets due to crystal field effects. Here, the estimated magnetic entropy for both PuNiSn and PuPtSn is reduced from the Rln(2) expected for a ground state doublet in the CEF. This may be due to similar energy scales between the Kondo effect and RKKY interation that both depend on the coupling parameter $J$, which describes the hybridization between conduction electrons and local moments. Magnetic order was observed in  Pu\textsubscript{2}Ni\textsubscript{3}Si\textsubscript{5} and Pu\textsubscript{2}Co\textsubscript{3}Si\textsubscript{5} with a reduced specific heat jump expected for a spin 1/2 system, reduced entropy expected for a ground state doublet, and a large linear specific heat coefficient both above and in the magnetically ordered state \cite{bauer_magnetic_2011}. This is thought to be due to Kondo and RKKY interactions of similar magnitude and is based on theoretical work which describes magnetic Kondo compounds with a doublet crystal field ground state using an $S=1/2$ resonant level model \cite{besnus_correlation_1992}. This work predicts a reduced specific heat jump at the antiferromagnetic ordering temperature as a function of $T_K$/$T_N$ and is in agreement with experimental evidence from various Ce and Yb compounds. In estimating the jumps in specific heat in Fig. 2, we find they are reduced from the expected 12.48 J mol\textsuperscript{-1} K\textsuperscript{-1} for both PuNiSn ($\Delta{C(11\ \mathrm{K}})\approx{3}$ J mol\textsuperscript{-1} Pu K) and PuPtSn ($\Delta{C(15\ \mathrm{K}})\approx{3.4}$ J mol\textsuperscript{-1} Pu K). This fact along with reduced magnetic entropy may indicate similar energy scales for Kondo and RKKY interactions in these materials.

The application of magnetic field has an extremely small effect on the peak in the specific heat of both compounds, as shown in Figs~\ref{fig:specific}(c) and (f). For PuNiSn, a field of $\mu_0H=14\ T$ pushes the transition to lower temperatures by only 0.25 K and slightly in magnitude. The peak for PuPtSn behaves very similarly in field and additionally has some field dependence at lower temperatures. The field dependence at low temperatures appears to be from a broad peak centered at approximately 3 K that is broadened with field as well as enhanced in temperature and magnitude. This is likely related to the small upturn seen in magnetic susceptibility at the same temperature. 

\subsection{Resistivity}

\begin{figure}[!ht]
\includegraphics[width=3.375in]{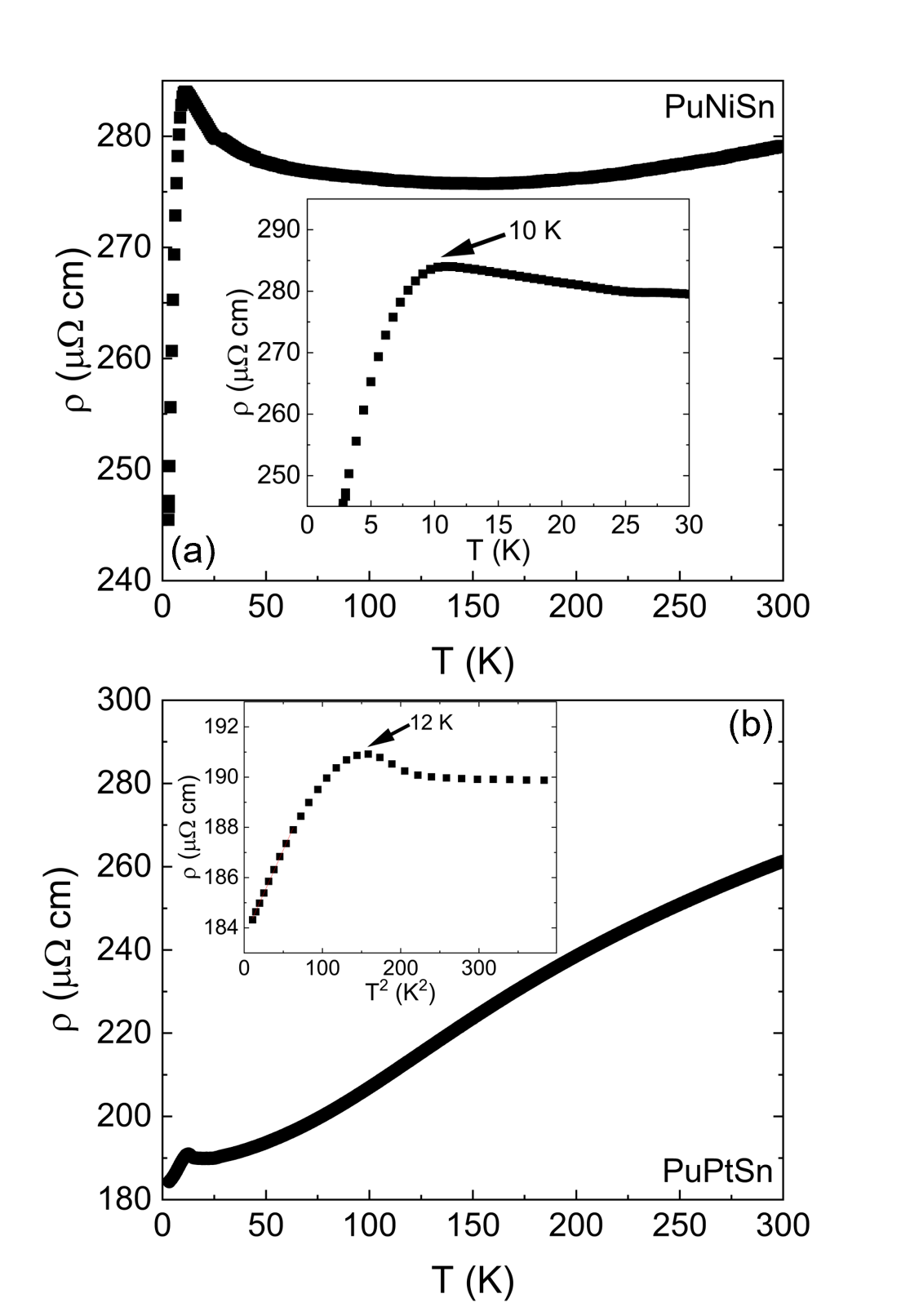}
\caption{\label{fig:res} Resistivity versus temperature for (a) PuNiSn and (b) PuPtSn. Insets: Manifestations of the AFM transitions show a sudden drop in resistivity in both compounds. }
\end{figure}

Figure~\ref{fig:res}(a) shows the temperature dependence of the resistivity for PuNiSn. PuNiSn shows typical Kondo behavior of a minimum in the resistivity followed by an increase as the temperature is lowered. There is a sharp decrease in the resistivity from a reduction in magnetic scattering as the moments order antiferromagnetically. As shown in the inset of Fig. \ref{fig:res}(a), the peak in resistivity coincides with the peak in magnetic susceptibility at $T_N=10$ K. NpNiSn, which also orders antiferromagnetically at 37 K, shows a similar overall temperature dependence \cite{javorsky_magnetic_1999}. However, the resistivity in the Np analogue has a negative slope from high temperatures down to the magnetic ordering temperature, which is thought to be from dynamic spin correlations rather than Kondo scattering. 

The resistivity for PuPtSn is plotted in Fig. \ref{fig:res}(b). It shows a very different temperature dependence than PuNiSn, where we find a monotonic decrease in resistivity down to low temperatures and a shallow minimum around 20 K, likely related to Kondo scattering or magnetic fluctuations in proximity to $T_N$. The maximum in resistivity at 12~K coincides well with the maximum in magnetic susceptibility. The temperature dependence of the resistivity is comparable to CePtSn, a Ce\textsuperscript{3+} ($J=5/2$) analogue that crystallizes in a TiNiSi-type structure \cite{alsmadi_magnetic_2011}. CePtSn is a Kondo compound with $T_K/T_N\approx{1}$, where crystal field splitting has an effect on the temperature dependence of the resistivity, as it was determined from inelastic neutron scattering that the first two excited states are $\Delta_1=273\ K$ and $\Delta_2=425\ K$ \cite{divis_determination_1994}. It is possible here that the thermal population of crystal field levels in PuPtSn also plays a role in the observed temperature dependence.

\subsection{Phonon Calculations}

\begin{figure*}[!ht]
\includegraphics[width = 6in]{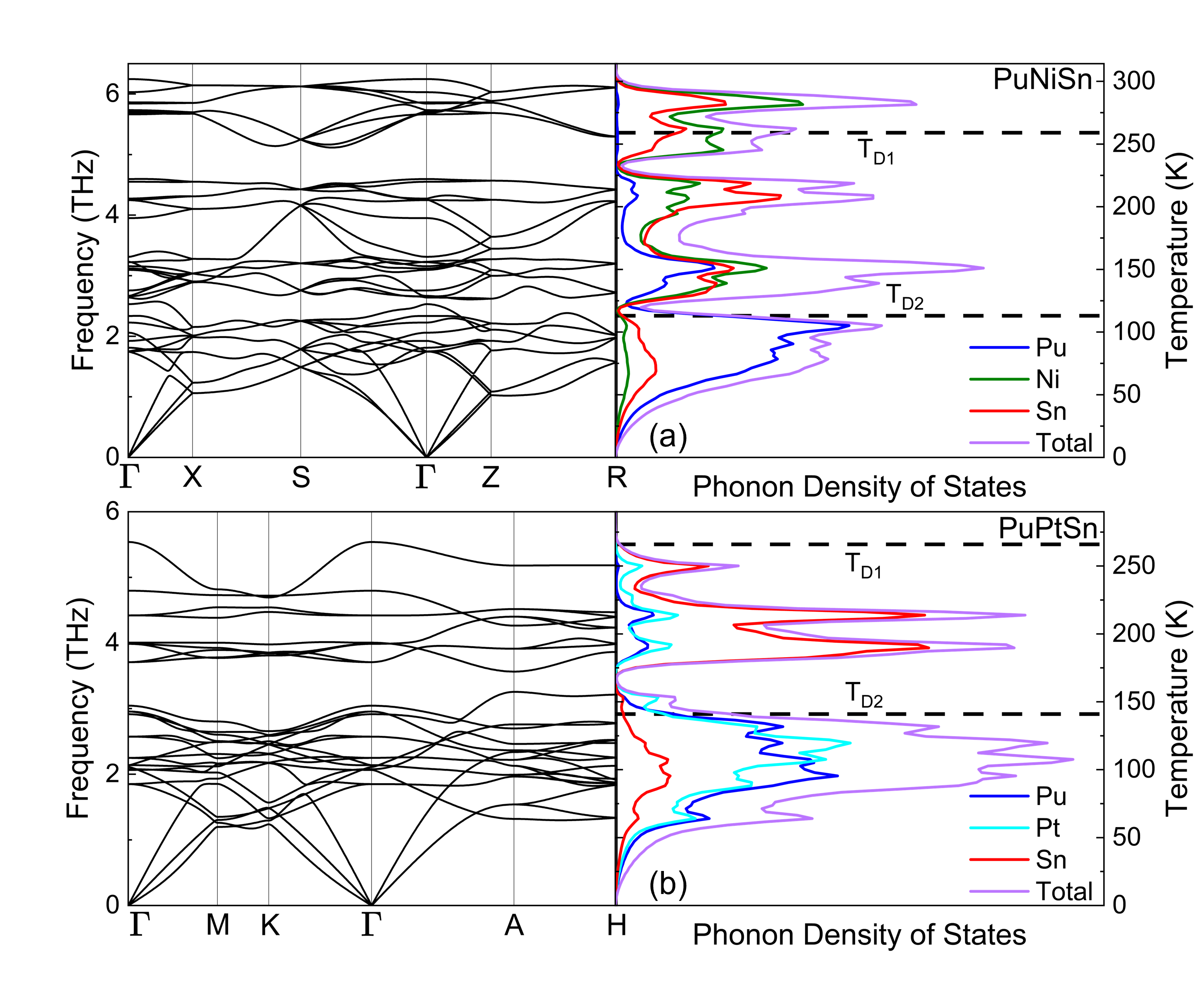}
\caption{\label{fig:DOS} Phonon dispersion (left) and corresponding density of states (right) for (a) PuNiSn and (b) PuPtSn. The dotted lines represent the experimental Debye temperatures from fits to the specific heat using equation (3).}
\end{figure*}

 The calculated phonon dispersion and partial phonon density of states (PPDOS) for PuNiSn and PuPtSn are presented in Fig.~\ref{fig:DOS}(a) and Fig.~\ref{fig:DOS}(b), respectively. For PuNiSn, the high frequency region is characterized by a gap, likely due to the large difference in mass of Pu compared to Ni and Sn, as relatively no phonon density comes from Pu above 5 THz. It has been previously shown that the phonon band gaps present in nitrides within the wurtzite structure can be increased by changing the cation-anion mass mismatch \cite{zhang_first-principle_2013}. Upon inspection of the PPDOS, the different vibrational properties of the constituent elements become apparent. The low frequency spectrum, especially below 2.5 THz, is mainly due to the vibration of Pu atoms, while the high frequency spectrum above 3.5 THz mainly comes from the lighter elements, Ni and Sn. There is however some Pu phonon density that coincides with the lighter elements in between these two regimes. This suggests the existence of phonon modes where all three elements vibrate together. In comparing this to fits to the lattice contribution of the specific heat we find that the two reported Debye temperatures which are show as dotted lines in Fig.~\ref{fig:DOS} correspond fairly well to the separate densities of the constituent elements. Our specific heat fit produces a Debye temperature of 113 K with $r=1$ which is consistent with Pu being responsible for most of the low frequency phonons.

 Similar arguments can be made for PuPtSn with some clear differences. We still find a gap in the phonon spectrum which clearly separates the Sn contributions in the higher frequencies from the Pu and Pt contributions in the lower frequencies. In this material, the transition element Pt (195.08 u) is much closer in mass to Pu (244.06 u) than Ni (58.69 u). Therefore we see in the phonon PDOS that most of the density of the lower frequencies are from Pu and Pt. This difference is also represented in the fit to the specific heat, where now $r=1.3$ and the smaller Debye temperature $T_{D2}$ has changed to 141~K as the Pu and Pt phonons occupy higher frequencies. Here, two Debye temperatures seem to capture the expected phonon density of states more accurately than for PuNiSn, as the dashed lines clearly show.


\section{Conclusion}

In conclusion, we have successfully used DFT calculations to predict the thermodynamic stability of two new Pu intermetallics, PuNiSn and PuPtSn. We have shown that exploratory synthesis of Pu compounds can be well informed through theoretical calculations, which offers a path forward when considering the difficulties associated with elemental Pu. We also show that our calculated phonon PDOS agree fairly well with modeling of the lattice specific heat. Using arc melting, we have grown these two compounds, confirmed the structures using PXRD, and subsequently characterized both materials using specific heat, magnetization, and resistivity measurements. Magnetization measurements show Curie-Weiss behavior at high temperatures followed by antiferromagnetic ordering of Pu\textsuperscript{3+} moments in both compounds at low temperatures. The low temperature anomalies in specific heat also support magnetic order, while a reduced magnetic entropy and jumps in the specific heat suggest a competition between the on-site Kondo and intersite RKKY interactions. Fits to the specific heat above the ordering temperature give an enhanced linear specific heat coefficient which is indicative of strong electron correlations in these materials. Resistivity measurements corroborate magnetic ordering in both PuNiSn and PuPtSn, as there is a prototypical drop in resistivity upon entering the ordered state from a reduction in magnetic scattering.  


\begin{acknowledgments}
All authors are grateful for support from LDRD's early career research catorgory under proposal number 20220538ECR. Further all authors are grateful to P.H. Tobash and L.M. Casados for useful discussions. E.D.B. was supported by the U.S. Department of Energy, Office of Basic Energy Sciences, Division of Materials Science and Engineering, under the project "Quantum Fluctuations in Narrow Band Systems." 
\end{acknowledgments}

\nocite{*}

\bibliography{bib_001}

\end{document}